%% file: RJwrapper.tex
\documentclass[a4paper]{report}
\usepackage[utf8]{inputenc}
\usepackage[T1]{fontenc}
\usepackage{RJournal}
\usepackage{amsmath,amssymb,array}
\usepackage{booktabs}

\begin{document}

\begin{article}
   \input{warriyarkv-almutiry-deardon}

\end{article}

\end{document}

%% file: warriyarkv-almutiry-deardon.tex
\title{ Individual-Level Modelling of Infectious Disease Data: \pkg{EpiILM}}
\author{by Vineetha Warriyar K. V.,  Waleed Almutiry and Rob Deardon}

\maketitle

\abstract{
In this article we introduce the R package \CRANpkg{EpiILM}, which provides tools for simulation from, and inference for, discrete-time individual-level models of infectious disease transmission proposed by \citet{deardon2010}. The inference is set in a Bayesian framework and is carried out via Metropolis-Hastings Markov chain Monte Carlo (MCMC). For its fast implementation, key functions are coded in Fortran. Both spatial and contact network models are implemented in the package and can be set in either susceptible-infected (SI) or susceptible-infected-removed (SIR) compartmental frameworks. Use of the package is demonstrated through examples involving both simulated and real data.}

\section{Introduction}
The task of modelling infectious disease transmission through a population poses a number of challenges. One challenge is that successfully modelling many, if not most, infectious disease systems requires accounting for complex heterogeneities within the population. These heterogeneities may be characterized by individual-level covariates, spatial clustering, or the existence of complex contact networks through which the disease may propagate. A second challenge is that there are inherent dependencies in infection (or event) times.

To model such scenarios, \citet{deardon2010} introduced a class of discrete time individual-level models (ILMs), 
fitting the models to data in a Bayesian Markov chain Monte Carlo (MCMC) framework. They applied spatial ILMs to 
the UK foot-and-mouth disease (FMD) epidemic of 2001, which accounted for farm-level covariates such as the number and type of animals on each farm. However, the ILM class also allows for the incorporation of contact networks through which disease can spread. Once fitted, such models can be used to predict the course of an epidemic \citep[e.g., ][]{OReilly2018} or test the effectiveness of various control strategies \citep[e.g., ][]{Tildesley2006} that can be imposed upon epidemics simulated from the fitted model. 

A third challenge when modelling disease systems is that very little software so far has been made available 
that allows for simulation from, and especially inference for, individual-level models of disease transmission. 
Most inference for such models is carried out in fast, low-level languages such as Fortran or variants of C, which 
makes it difficult for researchers (e.g., public health epidemiologists) without a strong background in computational 
statistics and programming to make use of the models. 

A number of R packages have recently been developed for modelling infectious disease 
systems (e.g., \CRANpkg{R0} \citep{r0}, \CRANpkg{EpiEstim} \citep{estim}, \CRANpkg{EpiModel} \citep{epimodel}, 
and \CRANpkg{epinet} \citep{epinet}). Most of these packages can be used to carry out epidemic simulation 
from  given models; in addition, \CRANpkg{R0} or \CRANpkg{EpiEstim}, for example, can be used to calculate the 
(basic) reproduction number under various scenarios. The \CRANpkg{EpiModel} package allows for the simulation 
of epidemics from stochastic models, primarily exponential-family random graph models (ERGMs), and provides
 tools for analyzing simulation output.  Functions for carrying out some limited forms of inference are 
 also provided. Another widely used package for monitoring and modelling infectious disease spread through 
 surveillance data is \CRANpkg{surveillance} \citep{surve}. This package provides for a highly flexible 
 modelling framework for such data. However, the package does not cover mechanistic, individual-level disease 
 transmission models such as those of  \citet{deardon2010}. 
 
Here, we detail a novel R statistical software package \CRANpkg{EpiILM} for simulating from, and carrying 
out Bayesian MCMC-based statistical inference for spatial and/or network-based models in the \citet{deardon2010} 
individual-level modelling framework. The package allows for the incorporation of individual-level susceptibility and  transmissibility 
covariates in models, provides various methods of summarizing epidemic data sets, and permits reasonably involved 
scenarios to be coded up by the user due to its setting in an R framework. The main functions, including for likelihood calculation are coded in Fortran in order to achieve the goal of agile implementation.  

The type of spatial and network-based transmission models that \CRANpkg{EpiILM} facilitates can be used to model a wide range of disease systems, as well as other transmissible processes. Human diseases such as influenza, measles or HIV, tend to be transmitted via interactions which can be captured by contact networks. For example, \cite {m2014} used a network representing whether two people shared the same household for modelling influenza spread in Hong Kong. Networks can also be used to characterize social or sexual relationships.

In the livestock industries, diseases are often transmitted from farm to farm via supply trucks or animal movements from farm to farm, or from farm to market. For example, ILM's were used by \cite {k2013} to model the spread of porcine reproductive and respiratory syndrome (PRRS) through Ontario swine farms via such mechanisms. Spatial mechanisms are also often important in livestock industries \citep[e.g., ][]{Jewell2009, deardon2010, k2013}, as well as for modelling crop diseases \citep[e.g.,][]{gyan}, since airborne spread is often a key factor.

Further, these types of models can also be used to model transmissible processes
other than infectious disease spread. For example, \cite{cook2007} used similar models to model the transmission of alien species through a landscape; specifically, giant hogweed in the UK. In addition, \cite{v2012} used spatial ILM's to model fire spread. They looked at fire spread under controlled conditions, but such models would likely be useful for modelling the spread of forest fires since important covariates such as vegetation-type could be incorporated into the models. 

Data from infectious disease systems are generally 'time-to-event', typically involving multiple states. However, standard survival models (e.g., \cite{cox}, \cite{survival}) or multi-state time-to-event models (e.g., see \cite{msm}) are not applicable here, because in an infectious disease system individual event times cannot be assumed independent even after conditioning on covariates. That is, my risk of contracting and infectious disease generally depends upon the disease state of other individuals in the population; this is not typically the case for most cancers, for example  to which more standard models can be applied.

The remainder of this paper is structured as follows: Section 2 explains the relevant models involved in the package; Section 3 
describes the contents of the package along with some illustrative examples; and Section 4 concludes the paper 
with a brief discussion on future development. 
\section{Model}
In our \CRANpkg{EpiILM} package, we consider two compartmental frameworks: susceptible-infectious (SI) and 
susceptible-infectious-removed (SIR). In the former framework, individuals begin in the susceptible state (S) 
and if/when infected become immediately infectious (I) and remain in that state indefinitely. In the latter framework, 
individuals once infected remain infectious for some time interval before entering the removed state (R). This final 
state might represent death, quarantine, or recovery accompanied by immunity. We consider discrete time scenarios so a complete epidemic history is represented by $ t = 1, 2, \dots, t_{end}$, where (typically) $t = 1$ is the time when the first infection is observed and $t_{end}$ is the time when the epidemic ends. Hence, for a given time point $t$, an individual $i$  belongs to one, and only one, of the sets $S(t)$ or $I(t)$ if the compartmental framework is SI, and $i$ belongs to one, and only one, of the 
sets $S(t)$, $I(t)$, or $R(t)$ if the compartmental framework is SIR.

Under either framework, the probability that a susceptible individual $i$ is infected at time point $t$ is given by $\P(i,t)$ as follows:

\begin{equation}
\P(i,t) =1- \exp\{-\Omega_S(i) \sum_{j \in I(t)}{\Omega_T(j) \kappa(ij)}-\varepsilon\}, 
\hspace{0.2cm} \Omega_S(i) >0, \hspace{0.1cm} \Omega_T(j) >0, \hspace{0.1cm} \varepsilon > 0
\label{eq1}
\end{equation}
where: $\Omega_S(i) $ is a susceptibility function that accommodates 
potential risk factors associated with susceptible individual $i$ contracting the disease; $\Omega_T(j) $ is a transmissibility function that accommodates 
potential risk factors associated with infectious individual $j$ contracting the disease; $\varepsilon$ is a sparks term which represents infections originating from outside the population being observed or some other unobserved infection mechanism; and $\kappa(i,j)$ is an infection kernel function that represents the shared risk factors between pairs of infectious and susceptible individuals.

The susceptibility function can incorporate any individual-level covariates  of interest, such as age, genetic factors, vaccination status, and so on. In Equation (\ref{eq1}), $\Omega_S(i)$ is treated as a linear function of the covariates, i.e., $\Omega_S(i) = \alpha_0 + \alpha_1 X_1(i) + \alpha_2 X_2 (i) + \dots  + 
\alpha_{n_s} X_{n_s} (i)$, where $X_1(i), \dots, X_{n_s} (i)$ denote  $n_s$ covariates associated with susceptible individual $i$, along with susceptibility parameters $\alpha_0,\dots,\alpha_{n_s} >0$.  Note that, if the model does not contain any susceptibility covariates then  $\Omega_S(i) = \alpha_0$ is used. In a similar way, the transmissibility function in Equation (\ref{eq1}) can incorporate any individual-level covariates of interest associated with 
infectious individual. $\Omega_T(j)$ is also treated as a linear 
function of the covariates, but without the intercept term, i.e., $\Omega_T(j) = \phi_1 X_1(j) + \phi_2 X_2 (j) + \dots  + \phi_{n_t} X_{n_t} (j)$, where $X_1(j), \dots, X_{n_t} (j)$ denote  the $n_t$ covariates associated with  infectious individual $j$, along with transmissibility parameters $\phi_1,\dots,\phi_{n_t} >0$.  Also
note that if the model does not contain any transmissibility covariates then  $\Omega_T(j)  = 1$ is used.

In this package, we also consider two broad types of ILM models based on the type of the kernel function $\kappa(i,j)$: spatial and network-based ILMs. In the spatial-based ILMs, the infection kernel function is represented by the power-law function as
\begin{equation*}
\kappa(ij) = d_{ij}^{-\beta},
\end{equation*}
where $\beta$ is the spatial parameter that accounts for the varying risk of transmitting disease over the Euclidean distance between individuals $i$ and $j$, $d_{ij}$.
Whereas in the network-based ILMs,  $\kappa(i,j)$ can be represented by one or more contact network matrices and is written as
\begin{equation*}
\kappa(ij) = \beta_1 \hspace{0.1cm} C^{(1)}_{ij}+\dots + \beta_n \hspace{0.1cm} C^{(n)}_{ij},
\end{equation*}
where $C^{(.)}_{ij}$ denotes the $(i,j)^{th}$ element of what we term the {contact matrix} of a given contact network; in graph theory this is more typically referred to as a (weighted) adjacency matrix. The  corresponding $\beta_{(.)}$'s represent the effect of each of the $n$ networks on transmission risk. In each contact network, each individual 
in the population is denoted by a node and is connected by lines or edges. These connections represent 
potential transmission routes through which disease can spread between individuals in the population. If 
the network is unweighted, the contact matrix is treated as binary (0 or 1). If the edges have weights assigned to them, 
then $C^{(.)}_{ij} \in R^+ $ or $C^{(.)}_{ij} \in [0,1]$ are typically used. These weights can be used to allow for different infection potential between different pairs of individuals. If the network is undirected, the contact 
matrix will be symmetric; if directed, it can be non-symmetric. Finally, the $C_{ii}$ (diagonal terms) are not used in 
the models and are typically set to $C_{ii}=0, \forall \hspace{0.05cm}i.$ 

Note that $\P(i,t)$ gives the probability that susceptible individual $i$ is infected at time point $t$, representing some interval in continuous time (e.g., a day or week),  but they actually become infectious at time $t+1$.

Following \citet{deardon2014}, the likelihood function for the ILMs (\ref{eq1}) is given by
\begin{equation}
f(S,I,R|\theta)=\prod_{t=1}^{t_{max}} f_t(S,I,R|\theta)
\end{equation}
where
\begin{equation}
f_t(S,I,R|\theta) =  \bigg[ \prod_{i \in I(t+1) \backslash I(t)} \P(i,t) 
\bigg] \bigg[ \prod_{i \in S(t+1)} 
(1-\P(i,t) ) \bigg]
\end{equation}
and where, $\theta$ is the vector of unknown parameters,  $I(t+1) \backslash I(t)$ denotes all new infections observed at $t+1$ in the infectious state at time $t$, and  $t_{max} \le t_{end}$ is the  last time point at which data are observed or being simulated.

\section{Contents of \CRANpkg{EpiILM}}
The \CRANpkg{EpiILM} package makes use of Fortran code that is called from within R. This package can be used to carry out  simulation of epidemics, calculate the basic reproduction number, plot various epidemic summary graphics, calculate the log-likelihood, and carry out Bayesian inference using Metropolis-Hastings MCMC for a given data set and model. The functions involved in the package are summarized in Table~\ref{table}. 
\begin{table}[h]
\centering
\begin{tabular}{p{3cm}  p{6cm} } 
 \toprule
Function & Output  \\ [1ex] 
\midrule
  
    \code{ epiBR0}  & Calculates the basic reproduction number for a specified SIR model  \\ [1ex] 
   
   \code{epidata}    & Simulates epidemic for the specified model type and parameters\\    [1ex] 
   
    \code{plot.epidata} & Produces spatial plots of epidemic progression over time as well as various epidemic curves of epidata object \\ [1ex]    
  
   \code{ epidic}    & Computes the deviance information criterion for a specified individual-level model  \\ [1ex] 
   
    \code{ epilike}  & Calculates the log-likelihood for the specified model and data set \\ [1ex] 
 
   \code{epimcmc}  & Runs an MCMC algorithm for the estimation of specified model parameters \\ [1ex] 
 
 \code{summary.mcmc} &  Produces the summary of epimcmc object\\ [1ex] 
 
 \code{plot.mcmc} &  Plots epimcmc object\\ [1ex] 
 
  \code{pred.epi} &  Computes posterior predictions for a specified epidemic model \\ [1ex] 
 
 \code{plot.pred.epi} &  Plot posterior predictions \\ [1ex]

 \bottomrule
\end{tabular}
\caption{Description of functions and their output in the \CRANpkg{EpiILM} package}
\label{table}
\end{table}

\subsection{Simulation of epidemics} 
The function \code{epidata()} allows the user to simulate epidemics under different 
models and scenarios. One can use the argument \code{type} to select the compartmental 
framework (SI or SIR) and population size through the argument \code{n}. 
If the compartmental framework is SIR, the infectious period is passed through the argument \code{infperiod}. Depending on whether a spatial or network model is being considered, the user can pass the arguments:  \code{x}, \code{y} for location and \code{contact} for contact networks. Users can also control the susceptibility function $\Omega_S(i)$ through the \code{Sformula} argument, with individual-level covariate information passable through this argument. If there 
is no covariate information, \code{Sformula} is null. An expression of the form 
\code{Sformula = $\sim$ model} is used to specify the covariate information, separated by \code{+} and \code{-} operators similar to the R generic function \code{formula()}. For example, $\Omega_S(i) = \alpha_0 + \alpha_1 X(i), \hspace{0.05cm} i = 1, \dots, n$ can be passed through the argument \code{Sformula} as \code{Sformula = $\sim$ 1 + X}. 
In a similar way, the user can control the transmissibility function $\Omega_T(i)$ through the \code{Tformula} argument. Note that, the \code{Tformula} must not include the intercept term to avoid model identifiability issues, i.e.,  for a model with one transmissibility covariate (\code{X}), the \code{Tformula}  becomes \code{Tformula = $\sim$ -1 + X}. 
The spatial (or network), susceptibility, transmissibility, and spark (if any) parameters are passed through arguments \code{beta}, \code{alpha}, \code{phi}, 
and \code{spark}, respectively.

The argument \code{tmin} helps to fix the initial infection time while generating an 
epidemic. By default, \code{tmin} is set as time $t=1$. We can also specify the initial 
infective or infectives using the argument \code{inftime}. For example, in a population 
of $10$ individuals, we could choose, say, the third individual to become infected at time point 1, using the option \code{inftime = c(0, 0, 1, 0, 0, 0, 0, 0, 0, 0)}. We could also infect more than one individual and they could be infected at different time points. This allows simulation from a model conditional on, say, data already observed, if we set the \code{tmin} option at the maximum value of \code{inftime}. 

The output of the function \code{epidata()} is formed as class of \code{epidata object}. This  \code{epidata object} contains a list that consist of  \code{type} (the compartmental framework), \code{XYcoordinates}(the XY coordinates of individual for spatial model) or \code{contact} (the contact network matrix for the network model), \code{inftime} (the infection times) and \code{remtime} (the removal times). Other functions such as \code{plot.epidata} and \code{epimcmc} involved in the package use this \code{object} class as an input argument.

\subsection{Descriptive analyses} 

We introduce an S3 method plot function to graphically summarize, and allow for a descriptive analyses of, epidemic data.
The function \code{plot.epidata()} illustrate the spread of the epidemic over time. One of the key input arguments (\code{x}) of this function has to be an \code{epidata object}. The other key argument \code{plottype}  has  two options: \code{curve}  and \code{spatial}.  Specifying the first option produces various epidemic curves,
while the latter show the epidemic propagation over time and space when the model is set to spatial-based. When the \code{plottype = curve}, an additional argument needs to be passed through the function through \code{curvetype}. This has  four options: \code{curvetype = "complete"} produces curves of the number of susceptible, infected, and removed individuals over time (when \code{type = "SIR''});\code{ "susceptible"} gives  a single curve for the susceptible individuals over time; \code{ "totalinfect"} gives the cumulative number of infected individuals over time; and \code{ "newinfect"} produces a curve of the number of newly infected individuals at each time point.

The plot functions \code{plot.epimcmc()} and \code{plot.pred.epi()} can be used to illustrate inference results (see Bayesian inference section). Detailed explanation is provided in the corresponding subsections.

\subsection{Example: spatial model} 
Suppose we want to simulate an epidemic from the model (1) using a spatial kernel with type SI, $\Omega_{s}(i) = \alpha$, and no transmissibility covariates, $\Omega_T(j) = 1$.  
Choosing the infectivity parameter $\alpha = 0.3$, spatial parameter $\beta = 5.0$, sparks parameter $\varepsilon = 0$, and $t_{max} = 15$, the model is given by:
\begin{equation}
\P(i,t) =1 - \exp\{-0.3 \sum_{j \in I(t)}{d_{ij}^{-5}}\}, \hspace{0.1cm} t=1,\dots, 15
\label{eq5}
\end{equation}
where $d_{ij}$ is the Euclidean distance between individuals $i$ and $j$, with their 
locations specified through  x and y. 

\noindent First, we install the \CRANpkg{EpiILM} package and call the library.
\begin{example}
R> install.packages("EpiILM")
R> library("EpiILM")
\end{example}
Then, let us simulate (x, y) coordinates uniformly across a $10 \times 10$ unit square.
\begin{example}
R> x <- runif(100, 0, 10)
R> y <- runif(100, 0, 10)
\end{example}
One could now use the following syntax to simulate an epidemic from spatial 
model (\ref{eq5}) and summarize the output as an epidemic curve and spatial plot.
\begin{example}
R> SI.dis <- epidata(type = "SI", n = 100, tmax = 15, sus.par = 0.3, beta = 5.0,
+                                x = x, y = y)
R> SI.dis$inftime
  [1]  0  0 10  4  7  0  0  0  4  5  0  0  9  2  6  6  3  0  0  0 11
 [22]  0  0  9 12  3  0 15  9  0  8  0  0  0 11  9  0  2  0  5  7  0
 [43]  2 15  7  0  0  5  0  0  0 14  8  2  0 15  9 10 10  4  1  5  4
 [64]  6 11  3  0  8  0  6  8  5 12  4  2 13  0  9  3  6  3  4  9 13
 [85]  0  0  0  0  0  0  0  0 10  0  9 11  9  9  0  0
\end{example}

\begin{figure}[h]
\centering
\includegraphics[width = 9cm, height = 9cm]{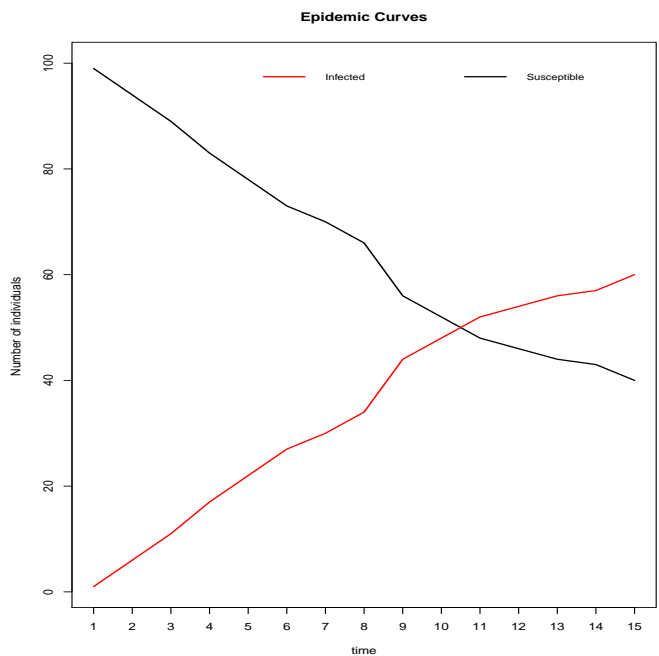}
  \caption{Epidemic curves of simulated epidemic from (\ref{eq5})  
  for 100 individuals at $t=1, \dots, 15$ }
  \label{si2}
\end{figure}

\begin{figure}[h]
\begin{minipage}[t]{ 7cm}
\includegraphics[width = 6.5 cm, height = 9cm]{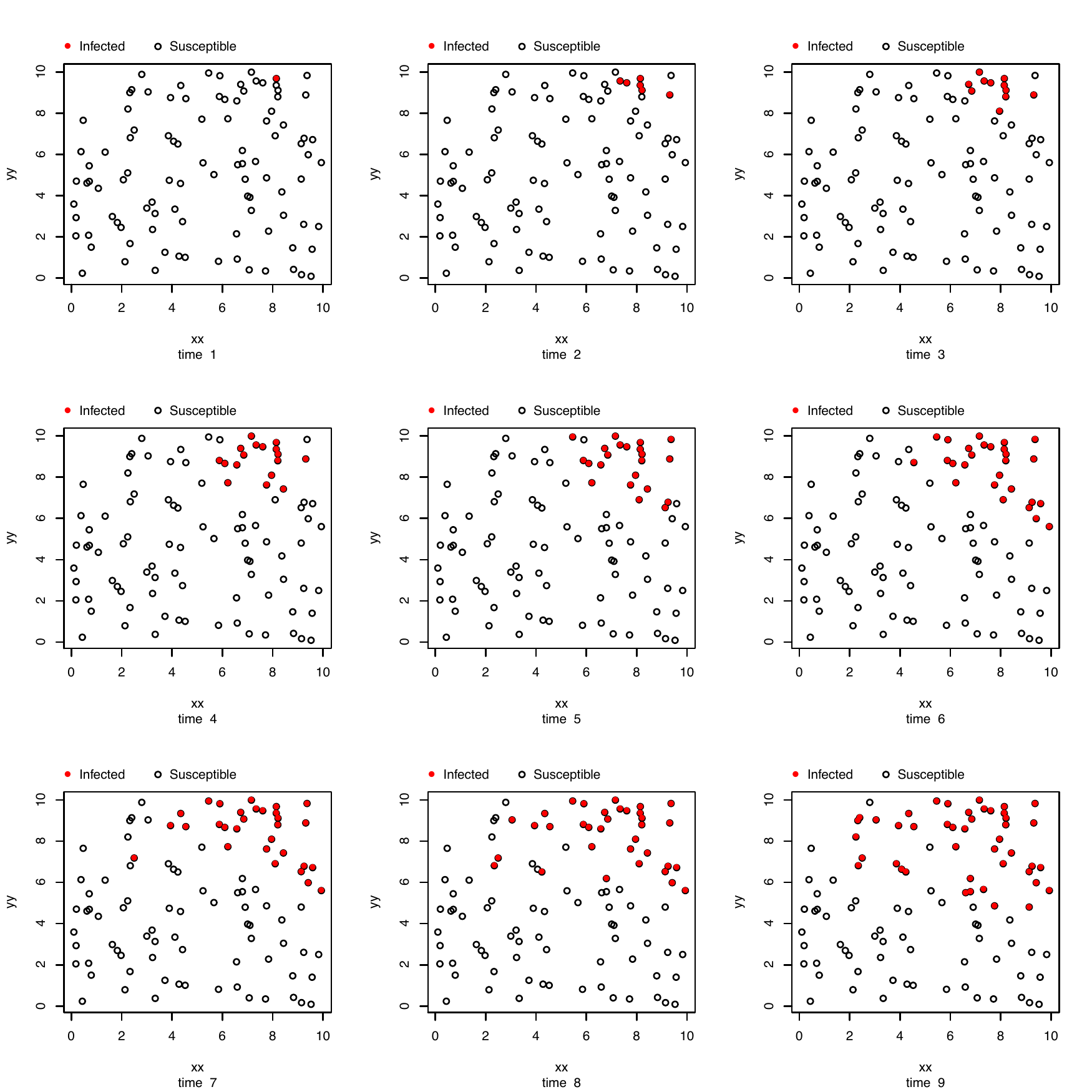}\\
 \end{minipage}
  \begin{minipage}[t]{7cm}
\includegraphics[width = 6.5 cm, height = 9cm]{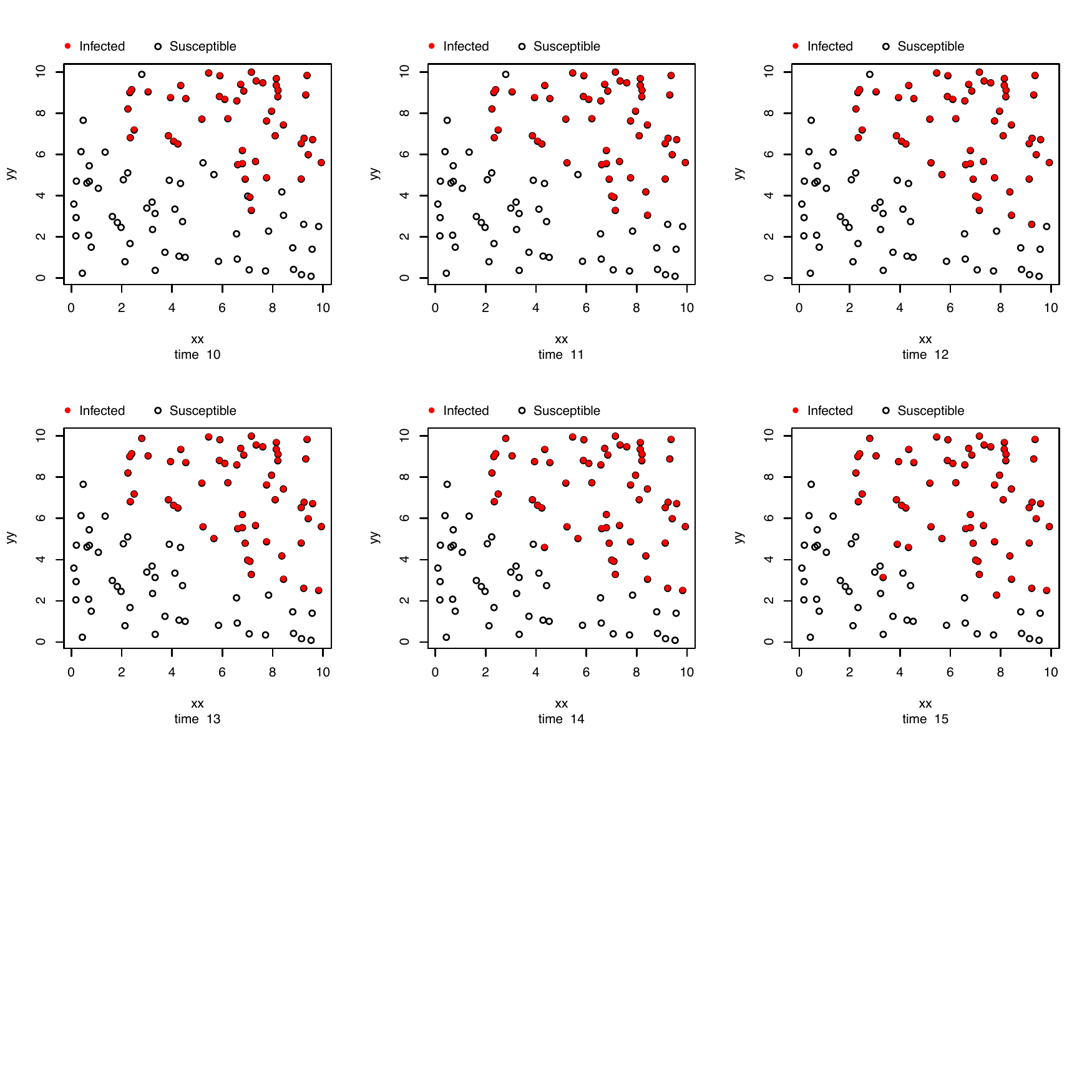}\\
 \end{minipage}
  \caption{Simulated epidemic from (\ref{eq5}) for 100 individuals, where 
  open circles represent susceptible individuals and filled red circles represent 
  infected individuals}
  \label{si1}
 \end{figure}

\noindent Here, the epidemic is generated across the uniformly distributed population of $100$ 
individuals with a default first infection time $t = 1$ and last observed time 
point of $t_{max} = 15$. The declaration of the spatial locations of individuals 
through \code {x} and \code {y} specifies that we are simulating from a spatial model.  
(See later for network-based models). 
The output  \code{SI.dis$\$$inftime} provides the times at which individuals enter the infectious state, with $0$ representing individuals who are still susceptible at time $t_{max}$. Figures~\ref{si2} and \ref{si1} 
 show the summary graphics for the simulated epidemic, which are produced using the 
 S3 method \code{plot.epidata()} as follows: 
\begin{example}
R> plot(SI.dis, plottype = "curve", curvetype ="complete")
R> plot(SI.dis, plottype = "spatial")
\end{example}
 \newpage

\subsection{Example: network model} 

\begin{figure}
\centering
\includegraphics[width = 9 cm, height = 9 cm]{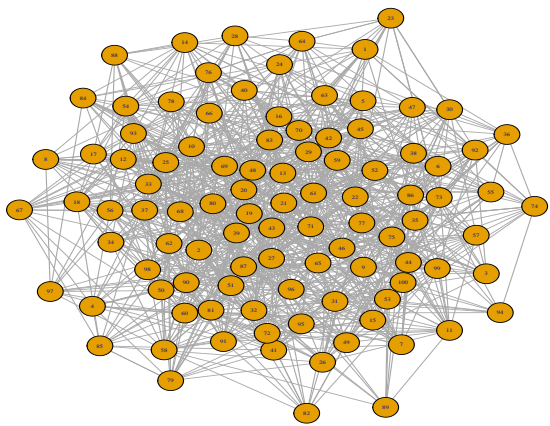}
 \vskip - 1cm    \caption{ Contact network generated for model (\ref{eq7}) }
  \label{net}
\end{figure}

\begin{figure}
\centering
\includegraphics[width = 10 cm, height =  10 cm]{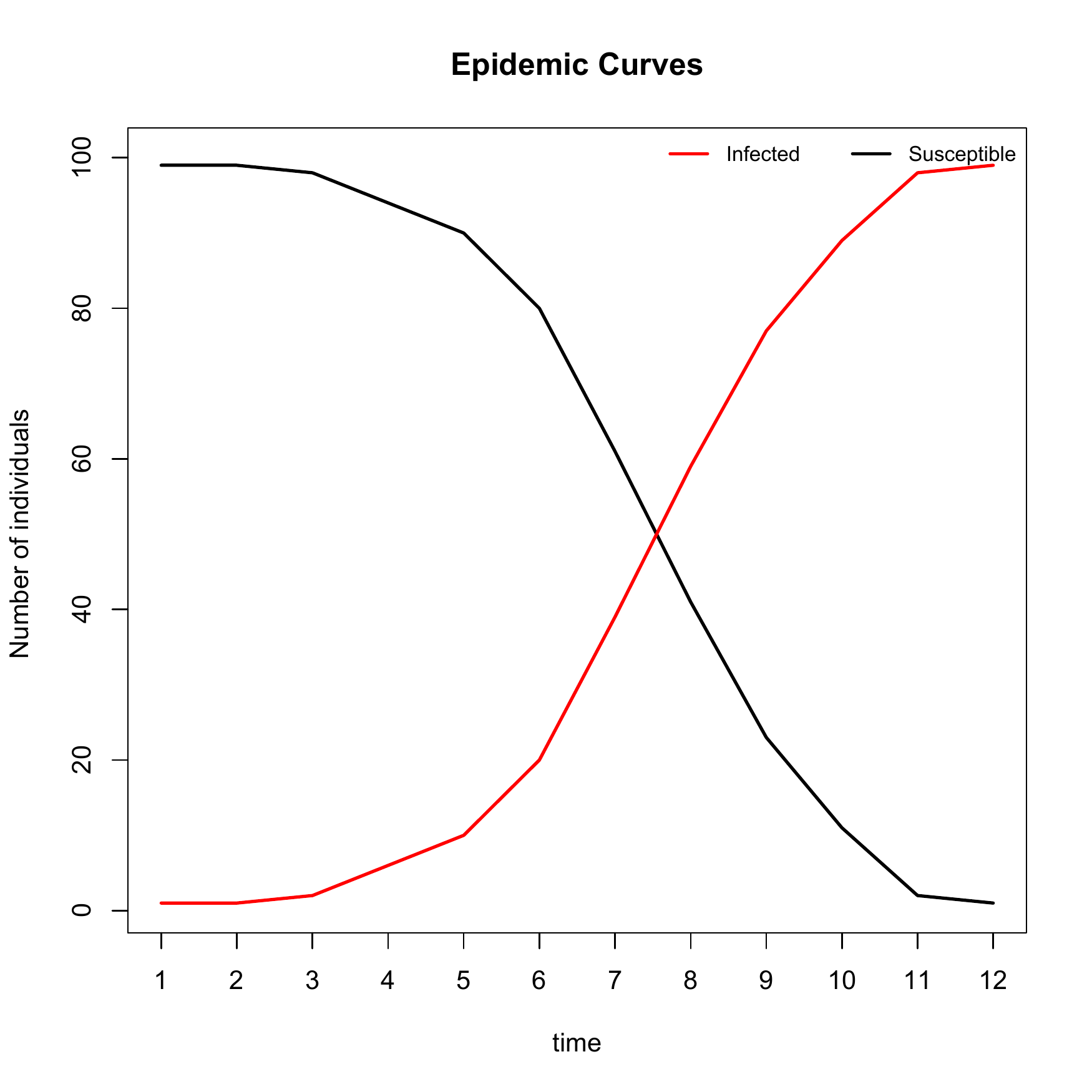}
  \caption{ Epidemic curve of contact network model (\ref{eq7}) for 100 individuals }
  \label{netepi}
\end{figure}

To illustrate simulation from a contact network-based model, we consider a disease system in which disease transmission can occur through a single, directed, binary network over a population of $n=100$ individuals. The elements in the contact matrix  represent the existence or non-existence of a directed connection through which disease can transmit between two individuals in the population. Each individual within the population is represented by a row and column within the matrix. Specifically, an element $C_{ij}$ in the contact matrix is given by:  

\begin{equation}
 C_{ij}=  \left\{
    \begin{array}{l}
      1 \hspace{0.5cm} \textrm{  if a directed  edge exists between $i$ and $j$ }\\
      0  \hspace{0.5cm} \textrm{ otherwise }
    \end{array}
  \right.
  \label{eq6}
\end{equation} 
We also consider the inclusion of a binary susceptibility covariate $Z$ in the model. This can be thought to 
represent, say, treatment or vaccination status. The infection model is then given by
\begin{equation}
\P(i,t) =1- \exp\{-(\alpha_0 + \alpha_1   Z_i) \sum_{j \in I(t)}{C_{ij}}\}, \hspace{0.1cm} t=1,\dots, t_{max}
\label{eq7}
\end{equation}
where $\alpha_0$ is the baseline susceptibility and $\alpha_1$ is the binary treatment effect.
The parameters in the model are set to be  $(\alpha_0, \alpha_1)= (0.1,0.05)$ and $\varepsilon=0$, and we also set $t_{max}=15$. We simulate a directed network using the following code:
\begin{example}
R> contact <- matrix(rbinom(10000, 1, 0.1), nrow = 100, ncol = 100)
R> diag(contact[, ]) <- 0
\end{example}
Various packages are available in R for network visualization, such as \CRANpkg{igraph}, \CRANpkg{ergm}, etc and as an example, we use the \CRANpkg{igraph} package for the network display as shown in Figure~\ref{net}.
\begin{example}
R> require("igraph")
R> net1 <- graph_from_adjacency_matrix(contact)
R> plot(net1, vertex.size = 10, vertex.label.cex = 0.5, edge.arrow.mode = "-")
\end{example}
The epidemic is generated using the function \code{epidata()}. The arguments \code{Sformula = $\sim$ Z}  and 
\code {sus.par = c(0.1, 0.05)}   define the  susceptibility function $0.1 + 0.05  Z_i$. As there is one contact network matrix in the model, the effect of the network $C$ is 
set to one by default. The use of the \code {contact} argument informs the package that we are dealing with a contact network-based model.
\newpage
\begin{example}
R> Z <- round(runif(100, 0, 2))
R> SI.contact <- epidata(type = "SI", Sformula = ~Z, n = 100, tmax = 15,
+           sus.par = c(0.1, 0.05),  contact = contact)
R> SI.contact$inftime
  [1]  0  9 11 10  9  5  8  9  8  9  9 10  1  7 12 11  7 11  8  4  8
 [22] 10  8 11  7  9 11  5  7  6  6  8  9  8  9  6  6  7  9  9  7  4
 [43]  7 10  8  3 10  9 10 11  6  7  9  6  5 11  4  7  8  8 10  8  8
 [64]  7  7  6 11  7  8  7  6  8  8 10  7  6 10 11  9  6 10  7  4  7
 [85]  8  7  9  9 10  9  8 10  5  8  7  7  8  9  9  8
\end{example}
Figure~\ref{netepi} shows the epidemic curve, obtained via the following code.
\begin{example}
R> plot(SI.contact, plottype = "curve", curvetype ="complete")
\end{example}

\subsection{Bayesian Inference}  
In \CRANpkg{EpiILM}, ILMs can be fitted to observed data within a Bayesian framework. A Metropolis-Hastings MCMC algorithm is provided which can be used to estimate the posterior distribution of the 
parameters. The function \code{epimcmc()} provides three choices for the marginal prior distribution of 
each parameter: the gamma, half-normal, and uniform distributions. The parameters are assumed 
to be {\it{a priori}} independent. The proposal used is a Gaussian random walk. Again, users can 
use the \code{Sformula} and \code{Tformula} arguments to specify any individual-level susceptibility and transmissibility covariates. Users can also control the number of MCMC simulations, initial values, and proposal 
variances of the parameters to be estimated. Note that in case of fixing one parameter and updating other parameters, users can do it by setting the proposal variance of fixed parameter to zero. This is usually the case to avoid identifiability issue when the model has both susceptibility and transmissibility covariates without intercept terms. Again, spatial/network, susceptibility parameters and transmissibility parameters are 
passed through arguments \code{beta}, \code{sus.par} and  \code{trans.par}, respectively. One can specify the 
spark parameter using the \code{spark} argument, but by default its value is $0$. 
\code{epimcmc()} can also call the adaptive MCMC method of inference facilitated by the \CRANpkg{adaptMCMC} package. Specifically, users can pass the argument  \code{adapt = TRUE} along with \code{acc.rate} to run the adapt MCMC algorithm.

\noindent We can use the S3 method functions \code{summary.epimcmc()} and \code{plot.epimcmc()} available in the package for output analysis and diagnostics. Both are dependent upon the \CRANpkg{coda} package. The argument \code{plottype} in the function \code{plot.epimcmc}, has two options to specify which samples are to be plotted: (1) "parameter" is used to produce trace plots (time series plots) of the posterior distributions of the model parameters, and (2) "loglik" to produce trace plots of the log likelihood values of the model parameter samples. Other options that are used in the \CRANpkg{coda} package can be used in the \code{plot.epimcmc()} as well; e.g.,  \code{start}, \code{end}, \code{thin}, \code{density},  etc.

\noindent Spatial or network-based ILMs can be fitted to data as shown in the following examples.

\subsubsection{Example: spatial model}

Suppose we are interested in modelling the spread of a highly transmissible disease through a series of farms, say $n=100,$ along with an assumption that the spatial locations of the farms and the number of animals on each farm are known. In this situation, it is reasonable to treat the farms themselves as individual units. If we treat the extent of infection from outside the observed population of farms as negligible ($\varepsilon = 0$), we can write the ILM model as  

\begin{equation}
\P(i,t) =1- \exp\{-(\alpha_0+\alpha_1 A(i)) \sum_{j \in I(t)}{d_{ij}^{-\beta}}\}, \hspace{0.1cm} t=1,\dots, t_{max}, 
\label{eq8}
\end{equation}

\noindent where the susceptibility covariate $A$ represents the number of animals on each farm,  $\alpha_0$ is the baseline susceptibility, $\alpha_1$ is the number of animals effect, and $\beta$ is the spatial parameter. Let us use the same (simulated) spatial locations from the previous spatial model example and set  the parameters $(\alpha_0, \alpha_1)= (0.2,0.1)$ and $\beta=5$.  We also set $t_{max}=50$. Considering an SI compartmental framework for this situation, the epidemic is simulated using the following command:

\begin{example}
R> A <- round(rexp(100,1/50))
R> SI.dis.cov <- epidata(type = "SI", n = 100, tmax = 50, x = x, y = y,
+                 Sformula = ~A, sus.par = c(0.2, 0.1), beta = 5)
\end{example}

\begin{figure}[h]
\centering
\includegraphics [width = 10cm, height = 10cm] {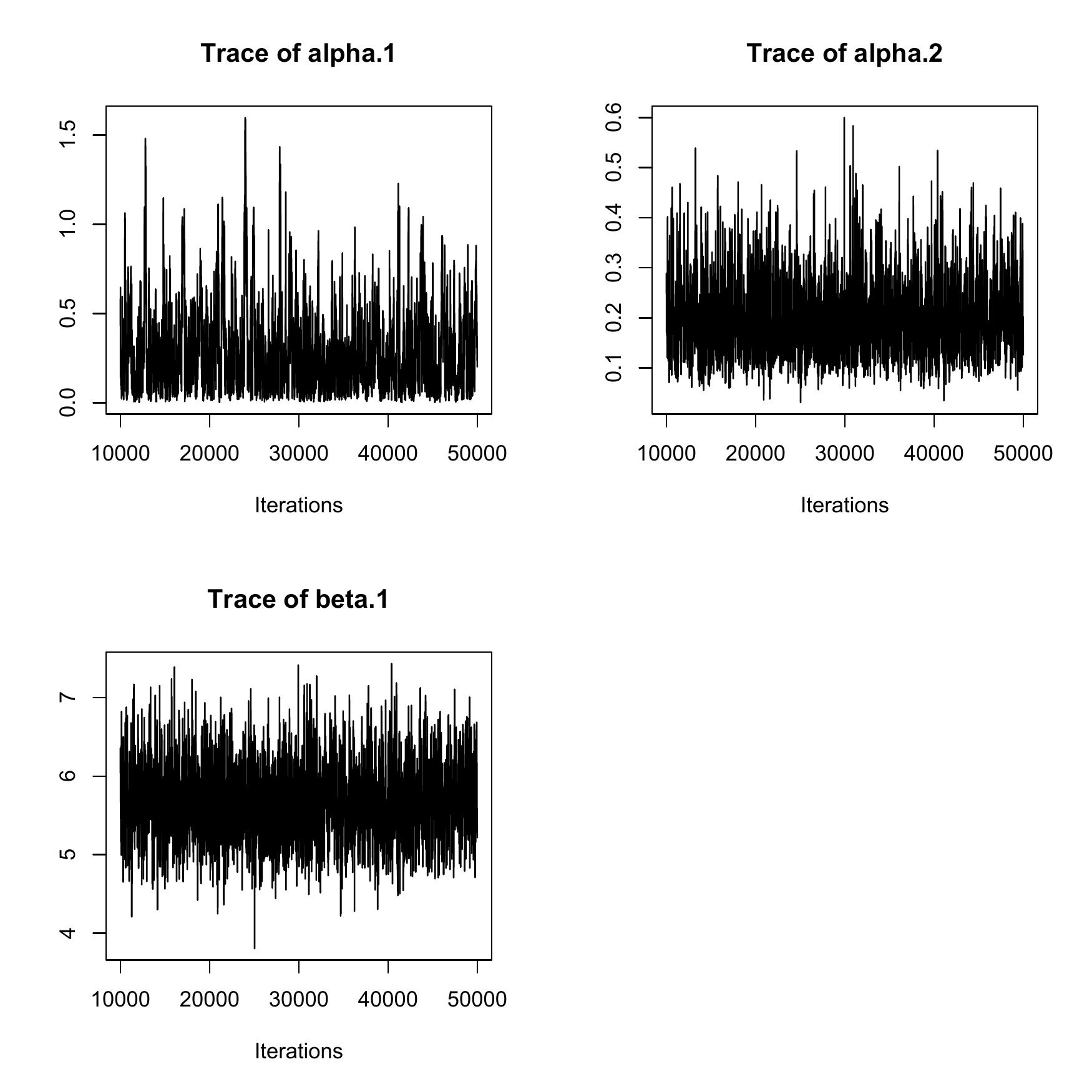}
\caption{ MCMC traceplot  for the estimation of model (~\ref{eq8}) parameters}
\label{si3}
\end{figure}

\noindent We can now refit the generating model to this simulated data and consider the posterior estimates of the model parameters. We can do this using the following code:

\begin{Sinput}
R> t_end <- max(SI.dis.cov$inftime)
R> unif_range <- matrix(c(0, 0, 10000, 10000), nrow = 2, ncol = 2)
R> mcmcout_Model7 <- epimcmc(SI.dis.cov, Sformula = ~A, tmax = t_end, niter = 50000, 
+                      sus.par.ini = c(0.001, 0.001), beta.ini = 0.01, 
+                      pro.sus.var = c(0.01, 0.01), pro.beta.var = 0.5, 
+                      prior.sus.dist = c("uniform","uniform"), prior.sus.par = unif_range, 
+                      prior.beta.dist = "uniform", prior.beta.par = c(0, 10000))
\end{Sinput}

\noindent where \code{niter} denotes the number of MCMC iterations and \code{sus.par.ini} and \code{beta.ini} are the initial values of the parameters to be estimated. The proposal variances for $(\alpha_0, \alpha_1)$ and $\beta$ are set to $(0.01, 0.01)$ and $0.5$, 
respectively (after tuning).  Vague uniform prior distributions are used for all three parameters, i.e., we choose $U(0, 10000)$ for $\alpha_0, \alpha_1$, and $\beta$. As the locations x and y are  specified, a spatial ILM is fitted rather than a network-based ILM. Note that the full data set to which the model is being fitted consists of the spatial locations (x, y) and the infection times. 
Figure~\ref{si3} displays the MCMC traceplot after 10000  burn-in using the command

\begin{Sin}   
R> plot(mcmcout_Model7, partype = "parameter", start = 10001, density = FALSE)
\end{Sin}

\noindent  The posterior means and $95\%$ credible intervals (CI) of the parameters, calculated as the $2.5\%$ and $97.5\%$ percentiles of 50000 MCMC draws after a burn-in of 10000 iterations has been removed, can be obtained using the following code:

\begin{Sout}   
R> summary(mcmcout_Model7, start = 10001)

Model: SI distance-based discrete-time ILM 
Method: Markov chain Monte Carlo (MCMC) 

Iterations = 10001:50000
Thinning interval = 1 
Number of chains = 1 
Sample size per chain = 40000 

1. Empirical mean and standard deviation for each variable,
   plus standard error of the mean:

          Mean      SD  Naive SE Time-series SE
alpha.1 0.2868 0.23516 0.0011758       0.012449
alpha.2 0.1997 0.07544 0.0003772       0.002393
beta.1  5.6790 0.48188 0.0024094       0.014631

2. Quantiles for each variable:

           2.5
alpha.1 0.02018 0.1141 0.2226 0.3917 0.9072
alpha.2 0.08680 0.1486 0.1875 0.2395 0.3769
beta.1  4.81361 5.3377 5.6704 5.9872 6.6588
\end{Sout}

\subsubsection{Example: network model} 

Now consider modelling the spread of an animal infectious disease through a series of farms ($n=500$) in a region. We once again consider individuals in the population to be the farms, rather than animals themselves, and a single binary contact network to represent connections between farms. We assume that there are two species of animals in farms that play an important role in spreading the disease; let us say, cows and sheep. Thus, we include the number of cows and sheep on the farm as susceptibility and transmissibility covariates in the model. Thus, the probability of susceptible farm $i$ to be infected at time $t$ becomes: 
\begin{equation}
\P(i,t) =1- \exp\{- (\alpha_{1}X_1(i) + \alpha_{2}X_2(i)) \sum_{j \in I(t)}{[\phi_{1}X_1(j) + \phi_{2}X_2(j)] C_{ij}}\}, \hspace{0.1cm} t=1,\dots, t_{max}
\label{eq10}
\end{equation}
\noindent with susceptibility parameters, ($\alpha_{1}, \alpha_{2}$), transmissibility parameters, ($\phi_{1}, \phi_{2}$), and $X_1$ and $X_2$ represent the number of sheep and cows in farms, respectively. Note, when we have a single network, there is no network parameter to estimate; this is to avoid problems of non-identifiability. 

\noindent We can generate such a directed contact network using the following commands:  
\begin{example}
R> n <- 500
R> contact <- matrix(rbinom(n*n, size = 1, prob = 0.1), nrow = n, ncol = n)
R> diag(contact) <- 0
\end{example}

\begin{figure}[h]
\centering
\includegraphics[width = 10cm, height = 8cm]{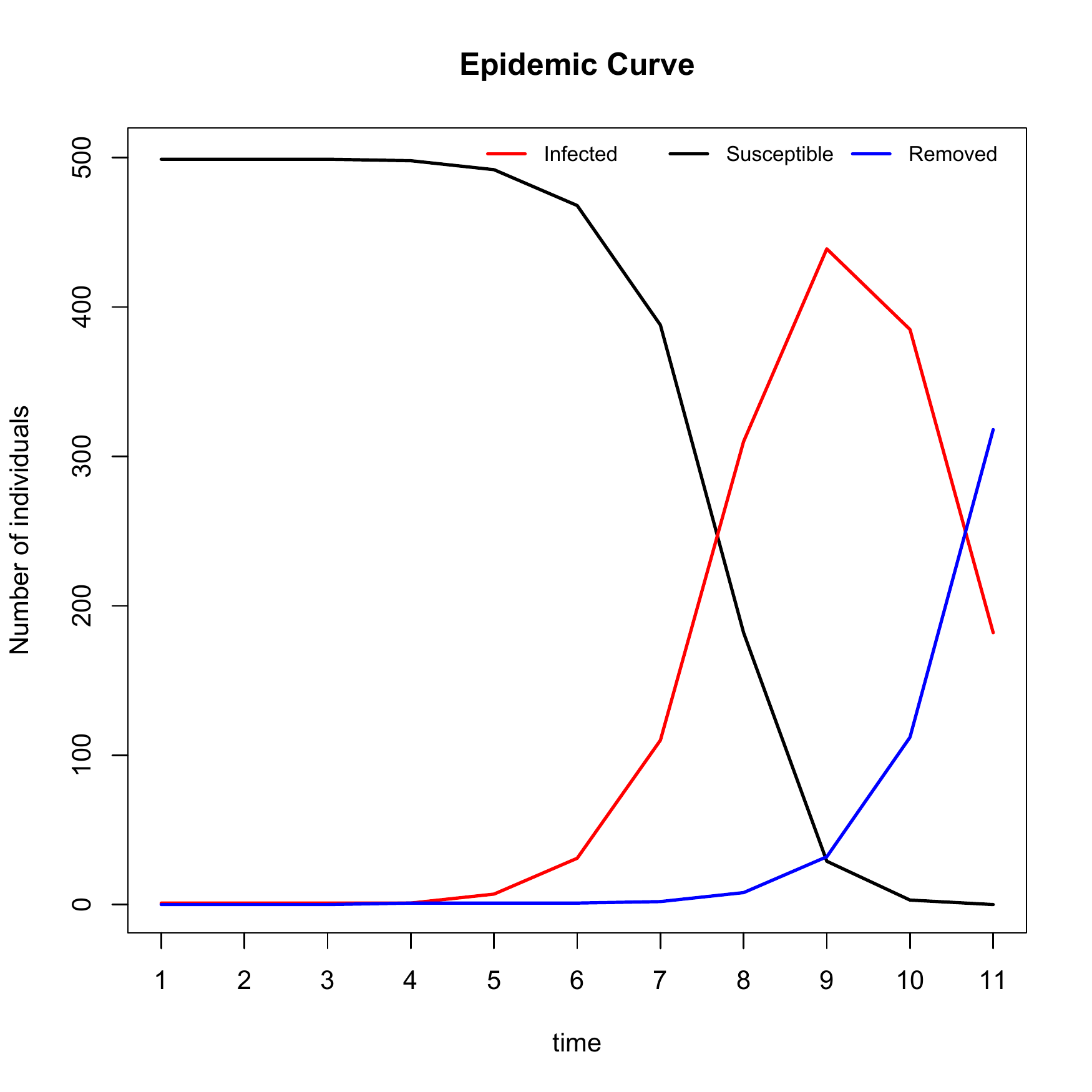}
  \caption{Epidemic curve of the contact network model (\ref{eq10}) for 500 farms}
  \label{SIR.net}
\end{figure}

\noindent We also sample the number of sheep and cows on each farm using the following code
\begin{example}
R> X1 <- round(rexp(n, 1/100))
R> X2 <- round(rgamma(n, 50, 0.5))
\end{example}
Assuming each infected farm to be infectious for an infectious period of 3 days and by setting $t_{max}=25$, $\alpha_{1}=0.003$, $\alpha_{2}=0.01$, $\phi_{1}=0.0003$, and $\phi_{2}=0.0002$, we can simulate the epidemic from the SIR network-based ILMs (\ref{eq10}) as follows and the epidemic curves are shown in Figure \ref{SIR.net}.
\begin{Sinput}
R> infp <- rep(3, n)
R>  SIR.net <- epidata(type = "SIR", n = 500, tmax = 15, 
+              sus.par = c(0.003, 0.01), trans.par = c(0.0003, 0.0002), 
+              contact = contact, infperiod = infp,
+              Sformula = ~ -1 + X1 + X2, Tformula = ~ -1 + X1 + X2)
R> plot(SIR.net, plottype = "curve", curvetype = "complete")
\end{Sinput}

\begin{figure}[h]
\centering
\includegraphics[width = 10cm, height = 9 cm]{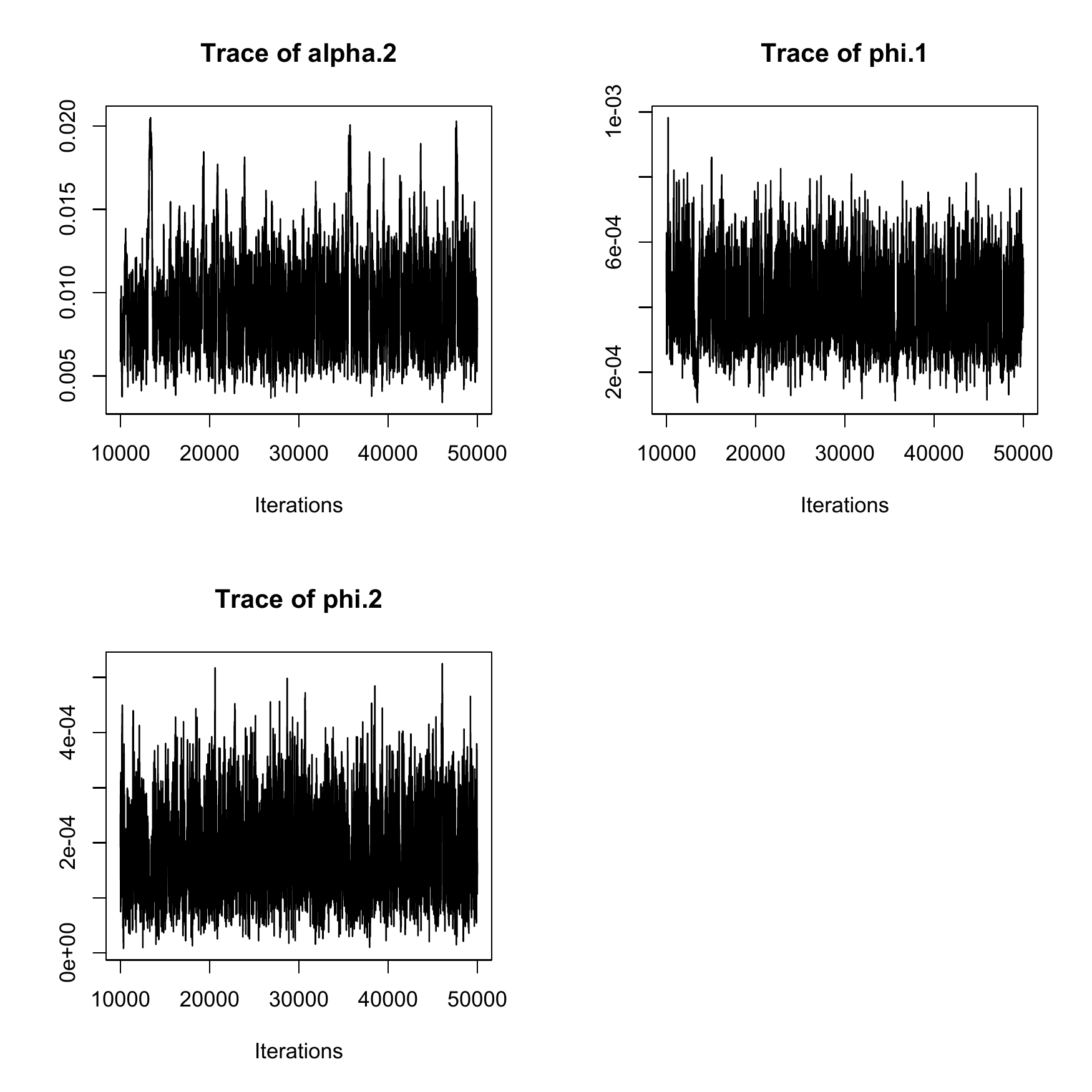}
  \caption{ MCMC traceplot for the estimation of model (\ref{eq10})  parameters.}
  \label{net2}
\end{figure}

\noindent To estimate the unknown parameters ($\alpha_{1}$, $\alpha_{2}$, $\phi_{1}$, $\phi_{2}$), we use the function \code{epimcmc()} assuming the event times (infection and removal times) and contact network are observed. Note that the specification of the contact argument means a network-based (rather than spatial) ILM will be assumed. We run the \code{epimcmc()} function to produce an MCMC chain of 50000 iterations assigning gamma prior distribution for the model parameters ($\alpha_{2}$, $\phi_{1}$, $\phi_{2}$) while fixing $\alpha_{1}$ to avoid the non-identifiablity issue in the model. This is done by setting the proposal variance of this parameter to zero in the \code{pro.sus.var} argument. To illustrate the use of adaptMCMC method, we set \code{adapt = TRUE} and set the acceptance rate as 0.5 via \code{acc.rate = 0.5}.  It should also be noted that we need to provide initial values and proposal variances of the parameters when we use the adaptive MCMC option in the \code{epimcmc()} function. The syntax is as follows: 
\begin{Sinput}
R> t_end <- max(SIR.net$inftime)
R> prior_par <- matrix(rep(1, 4), ncol = 2, nrow = 2)
R> mcmcout_SIR.net <- epimcmc(SIR.net, tmax = t_end, niter = 50000,
+              Sformula = ~-1 + X1 + X2, Tformula = ~-1 + X1 + X2,  
+              sus.par.ini = c(0.003, 0.001), trans.par.ini = c(0.01, 0.01),
+              pro.sus.var = c(0.0, 0.1), pro.trans.var = c(0.05, 0.05),
+              prior.sus.dist = c("gamma", "gamma"), prior.trans.dist = c("gamma", "gamma"), 
+              prior.sus.par = prior_par, prior.trans.par = prior_par,
+              adapt = TRUE, acc.rate = 0.5)
\end{Sinput}
\noindent Figure~\ref{net2}  shows the MCMC traceplot for the 50000 iterations.
The estimate of the posterior mean of the model parameters ($\alpha_{2}$, $\phi_{1}$, $\phi_{2}$) and their $95\%$ credible intervals, after 10000 iterations of burn-in have been removed are:
$\hat{\alpha_{2}} = 0.0094 \hspace {.1cm} (0.0051, 0.0168)$, $\hat{\phi_{1}} = 0.0004 \hspace {.1cm}(0.0002, 0.0007)$, and $\hat{\phi_{1}} = 0.0002 \hspace {.1cm} (0.00006, 0.0003)$.
\begin{Sinput}
R> summary(mcmcout_SIR.net, start = 10001)
\end{Sinput}
\begin{Sout}   
Model: SIR network-based discrete-time ILM 
Method: Markov chain Monte Carlo (MCMC) 

Iterations = 10001:50000
Thinning interval = 1 
Number of chains = 1 
Sample size per chain = 40000 

1. Empirical mean and standard deviation for each variable,
   plus standard error of the mean:

             Mean        SD  Naive SE Time-series SE
alpha.2 0.0093899 0.0030070 1.504e-05      1.866e-04
phi.1   0.0004123 0.0001218 6.089e-07      4.483e-06
phi.2   0.0001851 0.0000732 3.660e-07      1.911e-06

2. Quantiles for each variable:

             2.5
alpha.2 5.130e-03 0.0071826 0.0088076 0.0110550 0.0168404
phi.1   2.123e-04 0.0003226 0.0004020 0.0004879 0.0006800
phi.2   6.646e-05 0.0001322 0.0001763 0.0002299 0.0003463
\end{Sout}

\subsection{Case Study: Tomato spotted wilt virus (TSWV) data}
\begin{figure}[h]
\centering
\includegraphics[width = 10cm, height = 8cm]{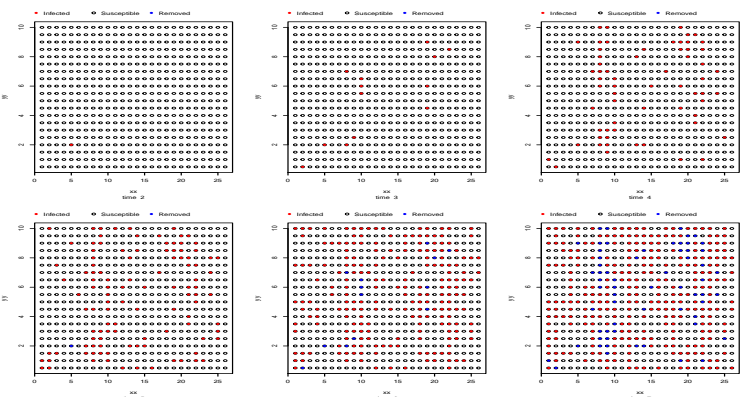}
\caption{ Epidemic dispersion of TSWV data across a grid of 520 individuals, where open circles
represent susceptible individuals, filled red circles represent infected individuals, and blue crosses represent the removed individuals.}
\label{tswv1}
\end{figure}

Here, we consider data from a field trial on TSWV as described in \citet{hugh}. The experiment was conducted on 520 pepper plants grown inside a greenhouse and the spread of the disease caused by TSWV was recorded at regular time intervals. Plants were placed in 26 rows spaced one metre apart, with 20 plants spaced half a metre apart in each row.  The experiment started on May 26, 1993 and lasted until August 16, 1993. During this period, assessments were made every 14 days. We set the initial infection time to be $t=2$ as the epidemic starts at time point 2 and the last observation made is set to be $ t = 7$. The TSWV data  in this package contain ID number, locations (x and y) and infectious and removal time of each individual. The data set was reconstructed from a spatial plot in \citet{hugh}. The infectious period for each tomato plant is assumed to span three time points (\citealp{brown2005}, \citealp{gyan}). Thus, an SIR model is fitted using the following code:
\begin{Sinput}
R> data(tswv) 
R> x <- tswv$x 
R> y <- tswv$y 
R> inftime <- tswv$inftime 
R> removaltime <- tswv$removaltime 
R> infperiod <- rep(3, length(x))  
\end{Sinput}
In order to see the spatial dispersion of the TSWV data, we use the function \code{epispatial()}. As no new infections occurred at the second time point, we display spatial plots from that second observation by setting \code {tmin = 2} (see Figure~\ref{tswv1}).  The required code is given by: 
\begin{Sinput}
R> epidat.tswv <- as.epidata(type = "SIR", n = 520, x = x, y = y, 
+                            inftime = inftime, infperiod = infperiod)
R> plot(epidat.tswv, plottype = "spatial", tmin = 2)
\end{Sinput} 

\begin{figure}[h]
\centering
\includegraphics [width = 10cm, height = 8cm]{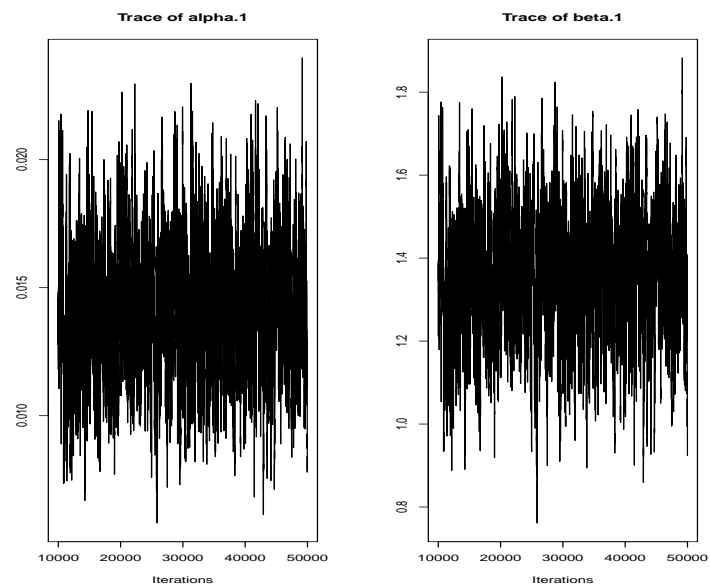}
\caption{ MCMC traceplots of the posterior samples for TSWV data}
\label{trace}
\end{figure}

The function \code{epimcmc()} is used to fit our model to the data. We ran $50000$ MCMC iterations. The computing time taken for the MCMC is about 10 min on a 16 GB MacBook Pro with a 2.9 GHz Intel Core i5 processor. Proposal variances are chosen to be $0.000005$ and $0.005$ for $\alpha$ and $\beta$, respectively (after tuning). We use vague independent marginal prior distributions for the parameters. Here, a gamma distribution with shape parameter $1$ and rate parameter $10^{-3}$ is used for both $\alpha$ and $\beta$. We choose to fit our model to data starting at the second time point because no infections occur between the first and second time points. Thus, we again set \code{tmin = 2}. The code for fitting our model is:
\begin{Sinput}
R> mcmc.tswv <- epimcmc(epidat.tswv, tmin = 2, tmax = 10,
+                niter = 50000, sus.par.ini = 0.0, beta.ini = 0.01,
+               pro.sus.var = 0.000005, pro.beta.var = 0.005, prior.sus.dist = "gamma",
+                prior.sus.par = c(1,10**(-3)), prior.beta.dist = "gamma", 
+                prior.beta.par = c(1,10**(-3)))			
\end{Sinput}
The  posterior mean estimates for this model are 
$\hat{\alpha}= 0.014$ and $\hat{\beta}= 1.351$. The $95\%$ credible interval of the posterior mean of $\alpha$ and $\beta$ are $(0.009, 0.0190)$ and  $(1.041, 1.629)$, respectively. Once again, these intervals are calculated as the $2.5\%$ and $97.5\%$ percentiles of $50000$ draws after 10000 iterations of burn-in have been removed.  The MCMC traceplots shown in Figure~\ref{trace}. 

\section{Conclusion}
This paper discusses the implementation of the R software package \CRANpkg{EpiILM}. Other than this package, there does not appear to be any R software that offers spatial and network-based individual-level modelling for infectious disease systems. Thus, this package will be helpful to many researchers and students in epidemiology as well as in statistics.
These models can be used to model disease systems of humans (e.g., \citet{m2014}), animals (e.g., \citet{k2013}), or plants (e.g., \citet{gyan}), as well as other transmission-based systems such as invasive species (e.g., \citet{cook2007}) or fire spread \citep{v2012}. \\
The \CRANpkg{EpiILM} package continues to exist as a work in progress. We hope to implement additional models and options in the future, that might  be useful for other researchers in their research and teaching. Such additions may include the incorporation of time-varying networks, covariates and/or spatial kernels (e.g.,  \citet{v2012}), models that allow for a joint spatial and network-based infection kernel, uncertainty in the times of transitions between disease states (e.g., \citet{malik2016}), unknown covariates (e.g., \citet{deeth}, and extensions to other compartmental frameworks such as SEIR and SIRS. Finally, we hope to extend the package to allow for the 
modelling of disease systems with multiple interacting strains or pathogens \citep{Razvan}.
\section{Acknowledgments}
Warriyar was funded by a University of Calgary Eyes High Postdoctoral Scholarship. Both Deardon, and equipment used to carry out this work, were funded by a Natural Sciences and Engineering Research Council of Canada (NSERC) Discovery Grant.

\bibliography{warriyarkv-almutiry-deardon}

\address{Vineetha Warriyar. K. V.\\
  Faculty of Veterinary Medicine\\
  University of Calgary\\
  Canada\\
  \email{vineethawarriyar.kod@ucalgary.ca}}

\address{Waleed Almutiry\\
  Department of Mathematics\\
  College of Science and Arts in Ar Rass\\
  Qassim University\\
  Saudi Arabia\\
  \email{wkmtierie@qu.edu.sa}}

\address{Rob Deardon\\
  Faculty of Veterinary Medicine and
  Department of Mathematics and Statistics\\
  University of Calgary\\
  Canada\\
  \email{robert.deardon@ucalgary.ca}}